\documentclass[journal]{IEEEtran}
\usepackage{ifpdf}
\usepackage{cite}

\usepackage{color}
\usepackage{amsthm}
\usepackage{amsmath,amssymb}
\usepackage{algorithmic}
\usepackage{array}
\usepackage{mdwmath}
\usepackage{mdwtab}
\usepackage{graphicx}
\usepackage{lipsum}
\usepackage{epstopdf }
\usepackage{algorithm,algorithmic}

\usepackage{enumerate}
\usepackage{eqparbox}
\usepackage{url}

\usepackage{subfig}
\usepackage{amsbsy}
\usepackage{multicol}
\usepackage{dblfloatfix}    
\usepackage{mathtools, cuted}

\usepackage{geometry}
\def\cI{\mathcal{I}}

\newcommand{\be}{\begin{equation}}
\newcommand{\ee}{\end{equation}}
\newcommand{\beq}{\begin{eqnarray}}
\newcommand{\eeq}{\end{eqnarray}}

  \newtheorem{theorem}{Theorem}
 
 \makeatletter
 \newif\ifnobrackets
 \renewcommand\@cite[2]{\ifnobrackets\else[\fi{#1\if@tempswa , #2\fi}\ifnobrackets\else]\fi\nobracketsfalse}
 \newcommand\nbcite{\nobracketstrue\cite}
 \makeatother
 
 \newcommand\numeq[1]%
 {\stackrel{\scriptscriptstyle(\mkern-1.5mu#1\mkern-1.5mu)}{=}}
 
\begin{document}
\pagestyle{empty}

\title{\vspace{-1mm}Performance Analysis of Blockchain Systems with Wireless Mobile Miners \vspace{-3mm}}
 \author{\IEEEauthorblockN{Gilsoo Lee, Jihong Park, Walid Saad, and Mehdi Bennis \vspace*{-12mm}}
}

\maketitle
\thispagestyle{empty}
\vspace{-0mm}
\begin{abstract}

In this paper, a novel framework that uses wireless mobile miners (MMs) for computation purposes in a blockchain system is proposed. 
In the introduced system, the blockchain ledger is located at the communication nodes (CNs), and the MMs associated with CNs process the blockchain's proof-of-work (PoW) computation to verify the originality of the data. 
The MM that is the first to finish its PoW will receive a reward by sending its computing result to the CNs that are connected to other MMs. 
In the considered scenario, a blockchain forking event occurs if the MM having the shortest PoW delay fails to be the first to update its computing result to other MMs. 
To enable such mobile operations for a blockchain with minimum forking events, it is imperative to maintain low-latency wireless communications between MMs and CNs. 
To analyze the sensitivity of the system to latency, the probability of occurrence of a forking event is theoretically derived.  
The system is then designed so as to compute  the forked block's PoW again to recover from a forking event. 
For this case, the average energy consumption of an MM is derived as a function of the system parameters such as the number of MMs and power consumed by the computing, transmission, and mobility processes of the MMs. 
Simulation results verify the analytical derivations and show that using a larger number of MMs  can reduce  the  energy  consumption by up to 94.5\% compared to a blockchain system with a single MM.
\end{abstract}

\IEEEpeerreviewmaketitle
\vspace{-9.5mm}
\section{Introduction}\vspace{-2mm}

As Internet of Things (IoT) connects physical systems to a network, and edge  computing enables to offload the computationally intensive tasks to the network edge, the emergence of blockchain is instrumental  in spearheading various verticals such as manufacturing, logistics, and autonomous systems \cite{dai2019blockchain, kim2019blockchained}. 
Blockchain applications can securely store data without a central trusted authority by leveraging distributed consensus mechanisms such as the so-called proof-of-work (PoW). 
{\color{black}Recently, blockchains have been adopted in the wireless domain  \cite{song2018blockchain, liang2017towards, suankaewmanee2018performance, xiong2018when, decker2013information}.} 
For instance, the authors in \cite{song2018blockchain} proposed a mobile blockchain network architecture that enables mobile devices to store  sensory data by using a blockchain. 
The work in \cite{liang2017towards} proposed a blockchain framework that uses drones to gather data that is then processed at cloud servers by carrying out  a mining process.  
{\color{black}The authors in \cite{suankaewmanee2018performance} developed a mobile blockchain application that  executes a mining process on a mobile device platform. }
Also, the authors in \cite{xiong2018when} applied a game-theoretic framework to solve a resource management problem in a mobile blockchain system where miners  offload their mining  tasks to the cloud. 
{\color{black}Moreover, the work in \cite{decker2013information} investigates the probability that a blockchain fork occurs for a given Bitcoin network protocol.} 

{\color{black}This prior art on mobile blockchains \cite{song2018blockchain, liang2017towards, suankaewmanee2018performance, xiong2018when, decker2013information} generally assumes that 
the miners store the ledger while updating their ledger by communicating with each other. 
However, if the miners cannot maintain a stable network connectivity due to the randomness of the wireless channel, or if the devices used to deploy a blockchain system do not have sufficient computing capabilities, a device cannot perform both mining and networking functions.} 
Therefore, it is more effective to consider a novel architecture for a blockchain system where mining and networking functions are migrated to two different types of nodes, respectively.  
{\color{black}Moreover, the prior works in \cite{suankaewmanee2018performance, xiong2018when, liang2017towards, decker2013information} do not account for the impact of the transmission latency at the wireless link of mobile devices on the blockchain performance.} 
Finally, the prior art mostly relies on isolated experimental results focused on simplistic use cases. 
{\color{black}However, a rigorous and generalized performance analysis of mobile blockchains in a wireless environment is needed to show how the parameters of a wireless blockchain system can affect its performance metrics such as forking (i.e., a situation when the node having the shortest PoW delay fails to be the first to update its computing result to other nodes) and device energy consumption.} 
Consequently, unlike the existing literature \cite{suankaewmanee2018performance, xiong2018when, liang2017towards} which investigates the use cases of mobile blockchains that store the ledger at the miners, our goal is to design and analyze a novel blockchain system using wireless mobile miners (MMs) such as drones or computationally capable moving nodes to process the mining computation while the ledgers are stored at the communication nodes (CNs) connected to MMs.

\begin{figure}[]
	\centering\vspace{0mm}
	\includegraphics[width=0.405\textwidth]{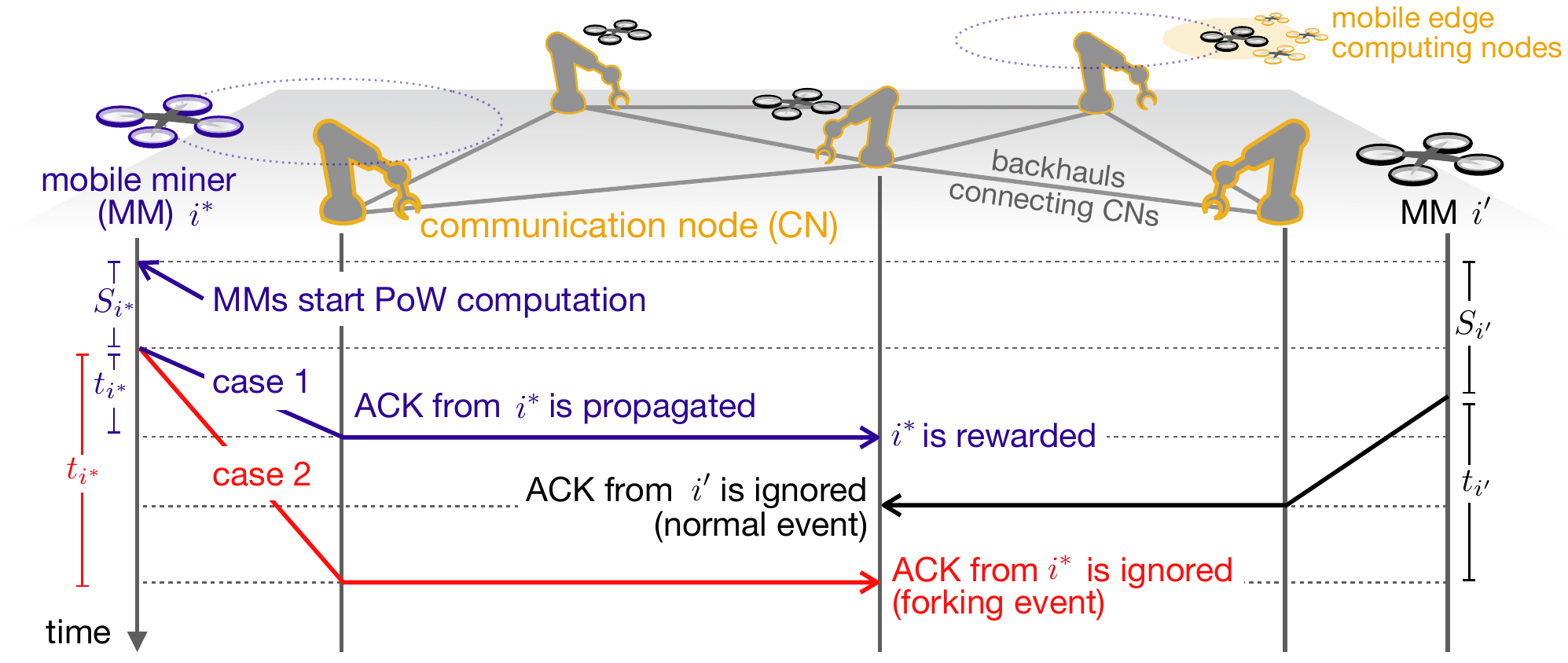}	\vspace{-3mm}
	\caption{\small \color{black}System model and timing diagram of normal and forking events.\vspace{-0.5mm}}
	\label{systemmodel}
	\vspace{-8mm}
\end{figure}

The main contribution of this paper is a novel, mobile blockchain system architecture and the performance analysis of the proposed system. 
In our architecture, each MM is connected to a CN via a wireless link, and the computing result of an MM is transmitted to other MMs through the backhaul network that interconnects the CNs. 
In such an architecture, forking events can occur when an MM propagates its computing result to other MMs, since the transmission latency between an MM and its associated CN can be large due to the wireless and mobile nature of the system. 
To this end, we derive an exact closed-form expression for the probability of occurrence of a forking event, as a function of the wireless network parameters such as the number of MMs and the MM power consumed by the MMs for computing, transmission, and mobility. 
We use the derived metric to find the average energy consumption required to process the PoW computation of a block. 
Our analytical result shows that the delay required for movement and the possibly high latency resulting from a wireless link can incur a forking event. 
Simulation results corroborate the analytical derivations and show that the energy consumption for PoW computation can be reduced by using a lower transmission power and decreasing movement of each MM. 

\newpage
\section{Mobile Blockchain Architecture}
\vspace{-2mm}

Consider a blockchain network consisting of  a set $\cI$ of $I$ MMs and CNs as shown in Fig.~\ref{systemmodel}. 
CNs can be seen as a fixed wireless network infrastructure such as base stations associated with MMs that can be computationally capable devices that can move such as industrial drones or ground vehicles gathering transaction data from other ground devices. 
{\color{black}When mining computation is executed by MMs, an MM can refer an independent computing node, and, also, an MM can refer the head node of a local computing cluster as shown in Fig.~\ref{systemmodel} \cite{lee2017mmtc}.}
We mainly focus on drone-type MMs due to their ability to flexibly deploy and move in nearly unconstrained locations\footnote{For example, when a smart factory consisting of industrial IoT devices owned by multiple operators calls for consensus, MMs can be wireless surveillance UAVs and warehouse robots, while CNs are wired-networked, fixed robotic assemblers and machines. 
Also in a vehicular network, MMs are vehicles, and CNs are road-side units and infrastructure \cite{dai2019blockchain}.}. 
{\color{black}However, our model can accommodate any other type of MMs. In Fig.~\ref{systemmodel}, MMs in remote locations are able to directly communicate each other. Therefore, each MM is associated with a different CN that is connected to a backhaul network.}
In the considered system, the ledger is located at the CNs while the MMs are used for computing. 
When transaction records are stored as blocks at the CNs, those blocks must be validated by PoW schemes so as to guarantee that the transaction in the block is original. 
Then, the CNs can delegate the PoW computation needed for this validation to the wireless MMs. 
Once each MM completes its PoW computation, an acknowledgment (ACK) message is sent from the MM to the associated CN, called the \emph{source CN}. 
The source CN propagates the reception of the ACK message to other CNs though the backhaul links among the CNs. 
{\color{black}We assume high-bandwidth, fiber backhaul links between CNs and, hence, the message propagation latency in  backhaul network will be negligible.} 

In a blockchain system, the reception order of the ACK messages from multiple MMs to the backhaul network should be identical to the order of completion of the PoW computation. 
If the ACK message sent earliest arrives to the CNs interconnected by the backhaul network later than other ACK messages, it will lead to a so-called \emph{forking event}. 
In a blockchain system, the MM that completes the PoW with the shortest delay will receive a unit of reward. 
However, when a forking event occurs, the MMs can no longer discern which MM completed the current PoW computation with the shortest delay. 
Therefore, if a forking event occurs, the MMs must recover from it by repeating the computation of the PoW of the forked block to decide which MM will be rewarded. 
This recovery from a forking event will clearly increase the total latency required to complete the PoW computation of a block.
Thus, \emph{the probability of occurrence of a forking event} is an important metric in a blockchain system that we will derive in Section~\ref{sec3}.
This metric will also allow us to analyze the average number of PoW computations and average MM energy consumption required to recover from a forking event.

\vspace{-5.5mm}
\subsection{System Model: Computation, Mobility, and Transmission}\vspace{-1mm}

In Fig.~\ref{systemmodel}, the computing latency $s_i$ and transmission latency $t_i$ of an MM $i\in\cI$ are the realization of  random variables $S_i$ and $T_i$, respectively. 
We assume that $S_i$ and $T_i$,  for all MMs $\forall i \in \cI$, will follow identical probability distributions. 
This is reasonable for the case in which all independent MMs are set to use the same computing power and wireless parameters. 
For notational simplicity, we use $S$ and $T$ to denote $S_i$ and $T_i$, respectively. 
Therefore, we will derive the probability distributions of $S$~and~$T$. \vspace{-0.3mm}

According to the PoW, all MMs start their PoW computation at the same time and keep executing the PoW computation until one of the MMs completes the computational task by finding the desired hash value \cite{suankaewmanee2018performance}. 
When an MM executes the computational task for the PoW of the current block, the time period needed to finish this PoW computation will be an exponential random variable $S$ whose distribution is 
$f_{S}(s) = \lambda_c e^{-\lambda_c s}$ 
where $\lambda_c=\lambda_0 P_{\textrm{c}}$ refers to the computing speed of an MM with $P_{\textrm{c}}$ being the power consumption for computation of an MM and $\lambda_0$ being a constant scaling factor. 

Once an MM finishes its PoW computation for the current block, the ACK message must be delivered to the associated CN so that other MMs can stop their current PoW computation. 
We derive the transmission latency when an MM transmits the ACK message to the associated CN through a wireless link. 
Under a Rayleigh fading channel, the small-scaling fading gain between an MM and the CN is a random variable $H$ with distribution $f_{H}(h)=\exp(-h)$ where the statistical average gain of the Rayleigh fading is unity. 
We assume that MMs move along a circular trajectory around the associated CN and, thus, they have a constant path loss $g$.
Then, the signal-to-noise-ratio (SNR) of any MM at its associated CN is the realization for the random variable~given~by:\vspace{-2.5mm}
\begin{equation}\label{gamma0}
	\Gamma_0 = {g H P_{\textrm{tx}}}/{\sigma_n^2}, 
\end{equation}\vspace{-6.5mm}\\
where $P_{\textrm{tx}}$ is the transmit power of an MM, and $\sigma_n^2$ is the noise power. 
Since $H$ is the only random variable in $\Gamma_0$, the distribution of the random variable $\Gamma_0$ will be $f_{\Gamma_0}(\gamma) = k_0 e^{- k_0 \gamma}$ 
where $k_0 = {\sigma_n^2}/({g P_{\textrm{tx}}})$. \vspace{-0.2mm}

\begin{figure*}[hb!] 
	\vspace{-5mm}
	\noindent\rule{\textwidth}{0.1pt}
	\vspace{-5mm}
	\begin{eqnarray}\label{nofork}
	p_{n}= e^{-\lambda_c({\color{black}I}-1)} \!\int_{0}^{\bar{t}}\!\!	\int_{0}^{\infty}  \left(\int_{0}^{t_{i^*}}\!\! e^{-\lambda_c(t_{i^*}-t)} f_T(t) d{\color{black}t} \!+\! \int_{t_{i^*}}^{\bar{t}} \!f_T(t) d{\color{black}t} \right)^{{\color{black}I}-1}  f_{S}(s_{i^*}) d{\color{black}s_{i^*}} f_T(t_{i^*}) d{\color{black}t_{i^*}}    
	\end{eqnarray}	
\end{figure*}

An MM transmits the ACK if the channel gain is higher than a threshold $\gamma_0$ that can be seen as the minimum SNR required to decode the transmitted data at the receiver. Hence, achieving an SNR higher than $\gamma_0$ is necessary to transmit the ACK data from an MM to the CN. 
In particular, each MM observes the SNR at any given location, and if the SNR is lower than $\gamma_0$, the MM moves to another location to obtain a better SNR.  
Hence, each MM will dynamically seek a location that yields an SNR higher than $\gamma_0$. 
The number of new location that an MM needs to visit can be given by $n+1, n \in \mathbb{Z^{\geq \textrm{$0$}}}$. 
At a given location, the probability that a certain MM achieves an SNR higher than $\gamma_0$ is $p_s=\textrm{Pr}(\Gamma_0 \geq \gamma_0) = 1- F_{\Gamma_0}(\gamma_0) = e^{-k_0 \gamma_0}$ where $F_{\Gamma_0}(\gamma)$  is the cumulative probability distribution of random variable $\Gamma_0$. 
{\color{black}Hence, the number of movements, $N$, can be modeled using a geometric distribution with the  probability mass function:}\vspace{-3.5mm}
\begin{eqnarray}\label{n}
f_N(n) = (1-p_s)^{n}p_s. 
\end{eqnarray}
\vspace{-7.0mm}\\
In order to change  the small-scale fading gain by moving from one location to another, an MM needs to move by a distance of $\lambda/2$ where $\lambda$ is the wavelength of the carrier frequency. 
The time period needed to move by a distance of $\lambda/2$ is given by $t_m= (\lambda/2)/v$ where $v$ is the speed of the MM. 
The power consumed\footnote{The power consumption needed to move a drone-type MM will be $P_{\textrm{m}}=P_H(v)+P_I(v)$ \cite{mozaffari2017mobile}. 
The closed-form equations of $P_H(v)$ and $P_I(v)$, are given in [\nbcite[Equations (57)-(59)]{mozaffari2017mobile}]. } to move an MM is $P_{\textrm{m}}$.
Therefore, the movement latency of the MM during $N$ movements becomes $T_m = t_m N$.

After finishing $N$ movements\footnote{\color{black}By setting zero velocity in our system model, a MM can be seen as a stationary mining node connected to backhaul network via a wireless link.}, the probability density function of SNR $\Gamma$, will be 
$f_{\Gamma}(\gamma) = g(\gamma)/(1-F_{\Gamma}(\gamma_0))$ 
where $g(\gamma) = f_{\Gamma_0}(\gamma) = k_0e^{-k_0 \gamma},$ if $\gamma_0 < \gamma$, and $g(\gamma) = 0$, otherwise. 
Therefore, the probability distribution of $\Gamma$ is rewritten~as
\vspace{-3mm}
\begin{eqnarray}
\scriptsize   f_{\Gamma}(\gamma) = 
  \begin{cases}
    k_0 e^{-k_0(\gamma-\gamma_0)},& \text{if } \gamma_0 < \gamma,\\
    0,              & \text{otherwise.}\nonumber
  \end{cases}
\end{eqnarray}
\vspace{-4mm}\\
The data rate of the MM is $R = B \log_2 \left(1+\Gamma \right)$ where $B$ is the bandwidth.  
The wireless transmission latency of the MM in the uplink will then be $T_{u} = {K}/{R}$ where $K$ is the size of the ACK message. 
Then, the probability density function of $T_u$ becomes $-f_{\Gamma}(v(t)) \frac{d}{d t}v(t)$ where $v(t)= 2^{\frac{K}{B t}}-1$, and, therefore, we have
\vspace{-2mm}
\begin{eqnarray}\label{tui}
\scriptsize   f_{T_u}(t) =
   		\begin{cases}
   				k_0 e^{-k_0 \left(2^{\frac{K}{B t}} -1- \gamma_0\right)} \frac{K \ln 2}{B t^2}  2^{\frac{K}{B t}}, &\hspace{-2mm}\text{if } 0 < t < \bar{t},\\
					0,              &\hspace{-2mm}\text{otherwise},\nonumber
			\end{cases}
\end{eqnarray} 
\vspace{-4mm}\\
where  $\bar{t} = K/(B \log_2(1+\gamma_0))$ is the largest wireless transmission latency in the uplink since the SNR is higher than $\gamma_0$. 
Thus, the total transmission latency including both movement and wireless transmission latencies becomes $T = T_m + T_u$. 

By using the random variables $S$, $T_m$, and $T_u$, the energy consumption of an MM in a single round of the PoW computation becomes a random variable given by: \vspace{-2mm}
\begin{equation}
	E=P_\textrm{c} S + P_m T_\textrm{m} + P_{\textrm{tx}} T_u.\nonumber
\end{equation} 
\vspace{-6mm}\\
Next, we analyze the performance of the proposed system by deriving the probability of no forking and the average energy consumption of an MM.

\begin{figure*}[t]
	\begin{multicols}{3}
		\hspace{-2mm}
		\includegraphics[width=1.1\linewidth]{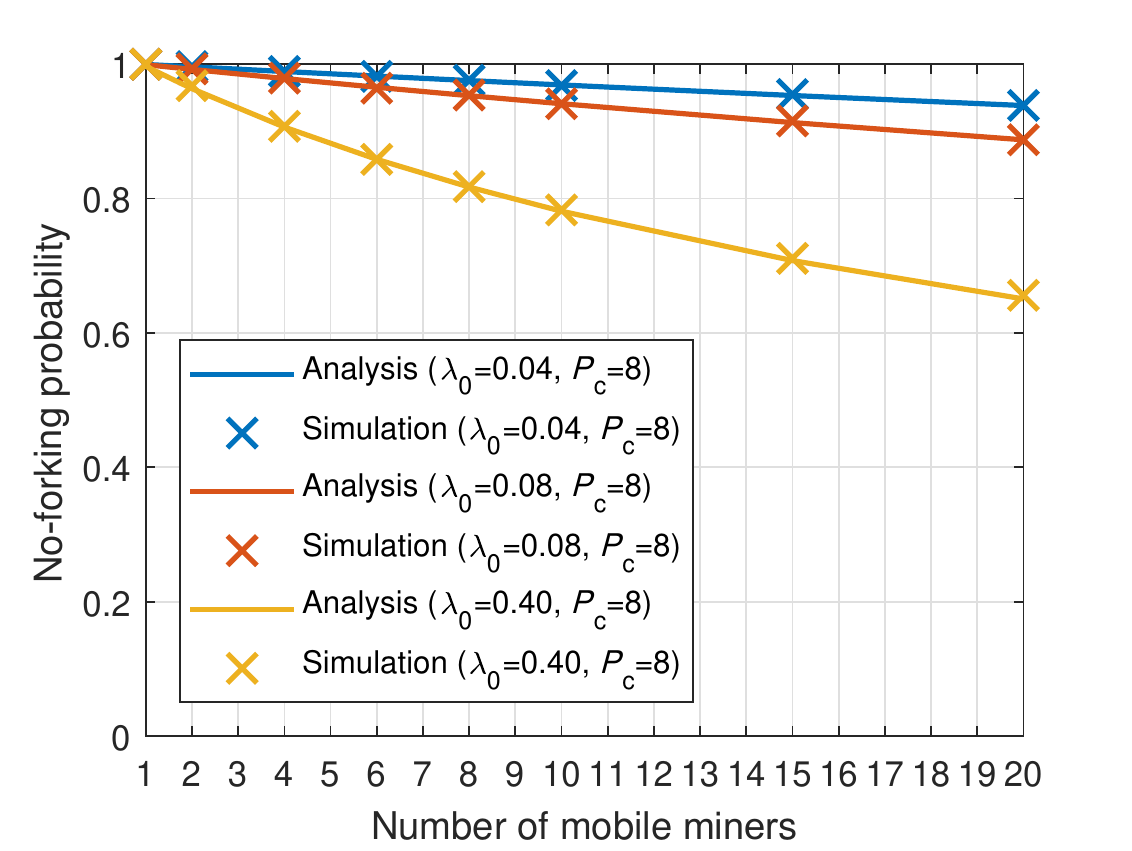}\par\caption{ \small Probability of a no forking event in one PoW computation.}\label{fig:prnof}
		\hspace{-4mm}
		\includegraphics[width=1.1\linewidth]{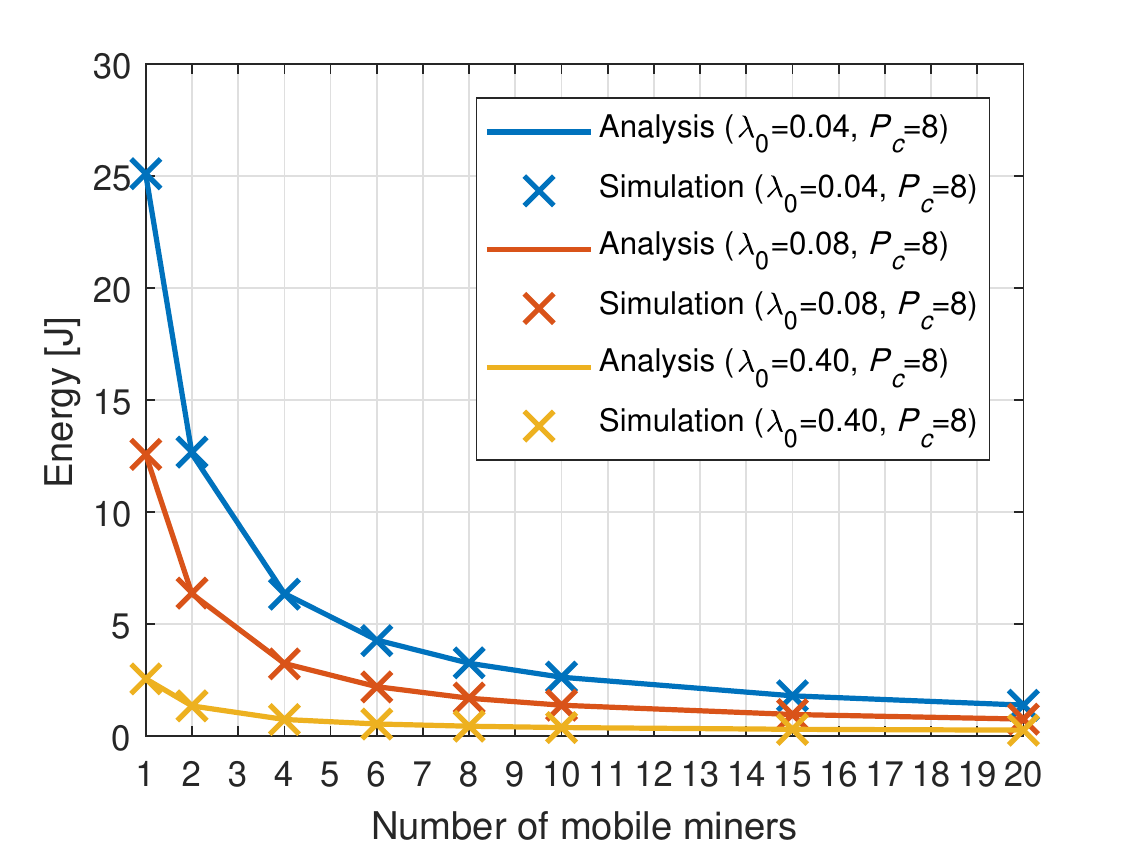}\par\caption{\small Average energy consumption of an MM required to compute the PoW of a block.}\label{fig:avge}
		\hspace{-6mm}
		\includegraphics[width=1\linewidth]{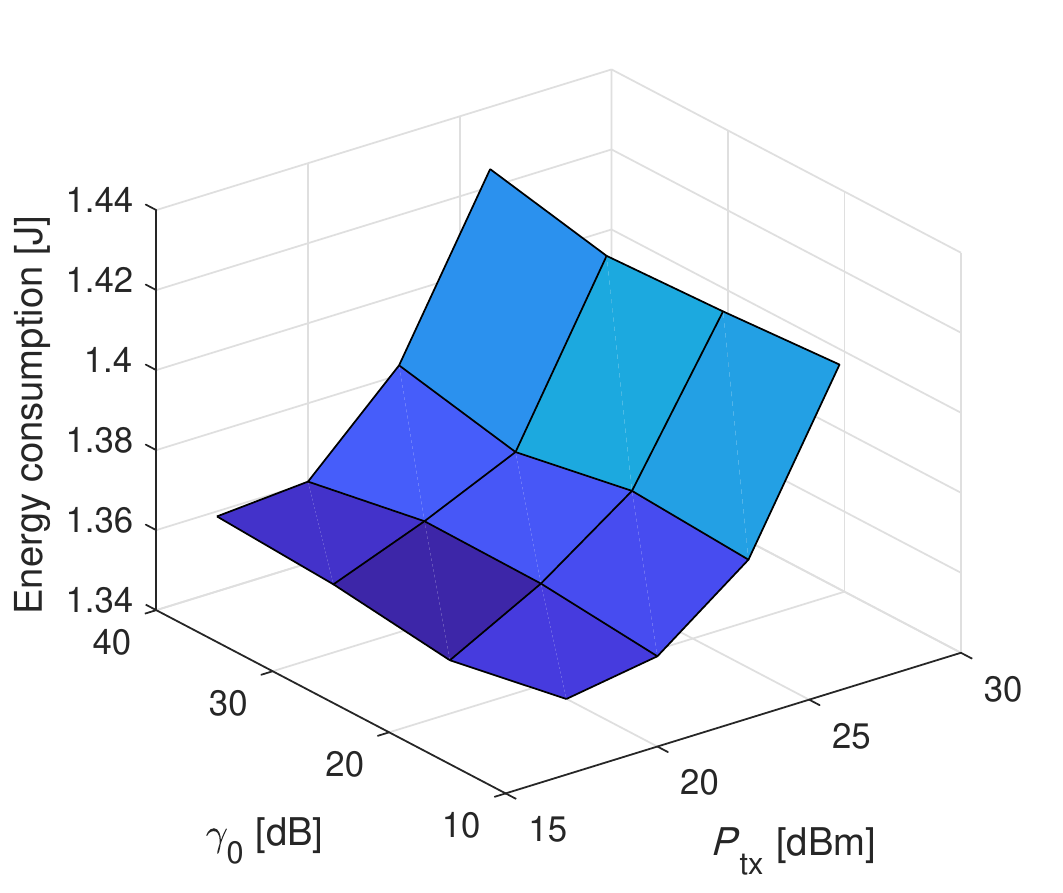}\par\caption{  \small Average energy consumption of an MM for different values of $P_{\textrm{tx}}$ and $\gamma_0$.}\label{fig:avge_tx_gamma}
	\end{multicols}\vspace{-9mm}
\end{figure*}

\vspace{-4mm}
\section{Average Energy Consumption Analysis}\label{sec3}
\vspace{-1mm}

We now analyze the performance of the proposed mobile blockchain system in terms of the occurrence of a forking event and the MM energy consumption.  
The occurrence of a forking event increases the energy consumption required to complete the current block's PoW due to the increment of the total computing and transmission latency used for the PoW re-computation. 
Therefore, the probability of having no forking event (called \emph{no-forking probability} hereinafter), that we derive next, is essential to derive the average MM energy consumption until one block is generated.

\vspace{-5mm}
\subsection{No-forking Probability}\vspace{-1mm}

For the PoW computation of a block, the MM that is the first to finish its PoW is indexed by $i^*$, i.e., $i^*=\textrm{argmin}_{i\in\cI} s_i$. 
Therefore, when $s_{i^*} < s_{i'}, \forall i'\in\cI\setminus\{i^*\}$, the ACK of MM~$i^*$ should arrive to the source CN so that the ACK information is propagated to all CNs via the backhaul network before any ACKs from other MMs arrive, i.e., $s_{i^*}+t_{i^*}<s_{i'}+t_{i'}$ as shown in Case~1 of Fig.~\ref{systemmodel}. 
However, the order of arrival of ACK messages from multiple MMs to the source CN can be different, i.e., $s_{i^*}+t_{i^*}>s_{i'}+t_{i'}$ as shown in Case~2 of Fig.~\ref{systemmodel}.
The change in the order of arrival happens when the transmission latency of MM~$i'$ is shorter than that of MM~$i^*$, i.e., $t_{i^*}>t_{i'}$, due to the different mobility patterns and the wireless transmission latency. 
Therefore, the ACK message of MM~$i'$ can arrive at the source CN earlier than the ACK message of MM~$i^*$, thus resulting in a forking event.  
We next derive the no-forking probability by calculating the probability that the ACK from MM~$i^*$ arrives earlier than any ACK from any other MM~$i' \in \cI\setminus \{i^*\}$. 

\vspace{-2mm}
\begin{theorem}\label{thm1}
	The no-forking probability is given by  \eqref{nofork}. 
\end{theorem}
\vspace{-4mm}
\begin{proof} See Appendix~\ref{proof_thm1}. 
\end{proof}\vspace{-2mm}
{\color{black}The no-forking probability $p_n$ is derived as a closed-form function of the wireless parameters such as the number of MMs, the MMs' transmission power, and the computing speed. 
If a forking event occurs, additional energy is needed for recovering from a forking event and, hence, we analyze the average MM energy consumption next. }

\vspace{-5mm}
\subsection{Average Energy Consumption}\vspace{-1mm}
Since a forking event incurs a PoW recomputation, the number of PoW computations needed to complete the current block's PoW follows a geometric distribution with mean of $1/p_n$. 
Also, MM~$i^*$ in each PoW computation will consume the average energy $\mathbb{E}[E]= P_{\textrm{c}} \mathbb{E}[S_{i^*}] + P_{\textrm{tx}} \mathbb{E}[T_u] + P_{\textrm{m}}\mathbb{E}[T_m] $. 
{\color{black}Our goal is to derive the average energy consumption of MM~$i^*$ in each PoW computation round until a block's PoW computation is completed without forking, i.e., $(1/{p_n})\mathbb{E}[E]$.} 

We outline how to derive  $\mathbb{E}[E]$ by finding the average latency of the computation, movement, and wireless transmission, i.e., $\mathbb{E}[S_{i^*}] $, $\mathbb{E}[T_u]$, and $\mathbb{E}[T_m]$, respectively. 
Since the shortest computing latency among all MMs is $S_{i^*}$, the complementary cumulative probability distribution (CCDF) of $S_{i^*}$ is given by $
\textrm{Pr}\left( S_{i^*} > z \right) =  \textrm{Pr}\left( \min_{i\in\cI}\left(S_i\right) > z \right) 
= \prod_{i=1}^{J} \textrm{Pr} \left(S_i > z \right) 
= \left(1 - \textrm{Pr}\left(S \leq z \right)\right)^{\color{black}I}. $
Therefore, the average computational latency of MM~$i^*$ is derived as $\mathbb{E}[S_{i^*}] = \int_{0}^{\infty} \left( 1-\textrm{Pr}(S \leq z) \right)^{\color{black}I} dz 
= \int_{0}^{\infty} e^{-\lambda_c {\color{black}I}  z}  dz 
= {1}/(\lambda_c {\color{black}I}) 
$.

The average latency due to mobility is given by $\mathbb{E}[T_\textrm{m}] = t_{m} (e^{(\gamma_0 \sigma_n^2)/(g P_{\textrm{tx}}) } -1)$ since the average of $N$ is 
$(1-p_s)/p_s 
= e^{(\gamma_0 \sigma_n^2)/(g P_{\textrm{tx}}) } -1$.  
Also, the average wireless transmission latency can be calculated by using the CCDF of the probability of $T_u$ given by
\vspace{-3mm}
\begin{eqnarray}
	\textrm{Pr}\left( T_u > z \right)
	&=& 
	\begin{cases}
	1- e^{-k_0(2^{\frac{K}{Bz}}-1-\gamma_0)},& \!\!\!\!\text{if } 0 \leq z \leq \bar{t}\\
	0,              & \!\!\!\!\text{otherwise. }
	\end{cases}\nonumber
\end{eqnarray}
\vspace{-4mm}\\
The average of the transmission latency is:
\vspace{-2mm}
\begin{equation}\label{Tuj}
	\mathbb{E}[T_u] = 	\int_{0}^{\bar{t}} 1- e^{-k_0(2^{\frac{K}{Bz}}-1-\gamma_0)} dz\nonumber
\end{equation}
\vspace{-4mm}\\
Hence, by combining $\mathbb{E}[S_{i^*}]$, $\mathbb{E}[T_u]$, and $\mathbb{E}[T_m]$, we can have a closed-form expression of $(1/{p_n})\mathbb{E}[E]$.

\vspace{-3mm}
\section{Simulation Results}
\vspace{-2mm}

For our simulations, we consider that an MM is associated with a CN at a distance of $50~\text{m}$, and the path loss gain $g$ is calculated by using a free space model due to air-to-air communications. 
The power spectral density of the  noise is $-174$~dBm/Hz, and the bandwidth is $180$~kHz. 
The ACK message size is set to $1$~Mbits. 
To model the computational speed of an MM, the power consumption for computation is set to $8$~W and the scaling factor $\lambda_0$ is $0.04$. 

{\color{black}Figs.~\ref{fig:prnof} and \ref{fig:avge} show the no-forking probability and the average energy consumption per MM.} 
First of all, Figs.~\ref{fig:prnof} and \ref{fig:avge} show that simulation and analysis results are matched. 
In Fig.~\ref{fig:prnof}, a forking event can occur with a high probability as the number of the MMs increases. 
This is due to the fact that, as more MM join in the PoW computing, the blockchain network is more likely to use an MM having a lower ACK reception period than that of MM $i^*$. 
Fig.~\ref{fig:avge} shows that the average energy consumption to complete the PoW computation decreases as the number of MMs increases because using more MMs reduces the PoW computation time period, thus decreasing the energy consumption. 
For example, using 20 MMs can reduce the energy consumption by up to 94.5~\% compared to using 1 MM in a blockchain system. 

{\color{black}Fig.~\ref{fig:avge_tx_gamma} shows the energy consumption required to complete the PoW computation.} 
The energy consumption of an MM increases with the SNR threshold $\gamma_0$ and the transmission power $P_{\textrm{tx}}$. 
This is because the energy consumption for mobility increases with $\gamma_0$. 
{\color{black}Also, the total energy consumption can increase with the transmission power because a high transmission power increases the forking probability, thus increasing the number of repeated PoW computations.}

\vspace{-3.5mm}
\section{Conclusion}
\vspace{-2.5mm}
{\color{black}In this paper, we have proposed a novel blockchain architecture with MMs. 
We have derived the no-forking probability and the average MM energy consumption.}
Simulation results have shown that the wireless transmission power, SNR threshold, and the number of MMs  significantly impact the energy consumption of the MMs. 
{\color{black}The analytical results serve as a nexus to minimize the forking probability over wireless and device energy consumption as a future work.}
\vspace{-5mm}

\appendices
\section{Proof of Theorem~\ref{thm1}}\label{proof_thm1}\vspace{-1mm}

\begin{proof}\vspace{-1mm}
	
	As a first step, suppose that the shortest computing latency is $s_{i^*}$, and that MM~$i^*$ has a transmission latency $t_{i^*}$. 
	Given $s_{i^*}$ and $t_{i^*}$, the probability that all MMs other than MM~$i^*$, do not incur a forking event becomes:
	\vspace{-3mm}
	\begin{eqnarray}
	&&\hspace*{-10mm}	\textrm{Pr}\bigg(  \bigcap_{i' \in \cI \setminus \{i^*\}} \!\!\!\!\! s_{i^*}+t_{i^*} < S_{i'} + T_{i'} | S_{i^*}=s_{i^*}, T_{i^*}=t_{i^*}, s_{i^*} < S_{i'}\!\!\bigg) \nonumber\\
	&\numeq{a}&\hspace*{-5mm}\prod_{i'\in \cI\setminus\{i^*\}} \!\!\!\!\!	\textrm{Pr}\bigg( s_{i^*}\!+\!t_{i^*} \!\!<\! S_{i'} \!+\! T_{i'} | S_{i^*}\!=\!s_{i^*}, T_{i^*}\!=\!t_{i^*}, s_{i^*} \!\!<\! S_{i'}\bigg)\nonumber\\
	&\numeq{b}&\!\!\!\! \bigg(\textrm{Pr}\left( s_{i^*}\!+\!t_{i^*} \!\!<\! S \!+\! T | S_{i^*}\!=\!s_{i^*}, T_{i^*}\!=\!t_{i^*}, s_{i^*} \!\!<\! S \right)\bigg)^{{\color{black}I}-1}\nonumber
		\end{eqnarray}
	\begin{eqnarray}
	\numeq{c}&\!\!\!\!\bigg( \int_{0}^{\bar{t}}\! \int_{\max(s_{i^*}, s_{i^*}+t_{i^*}-t)}^{\infty} f_{S}(s|s_{i^*}<S)  d{\color{black}s} f_T(t) d{\color{black}t}  \bigg)^{{\color{black}I}-1}	&\nonumber\\
	=&\!\!\!\! \bigg(e^{-\lambda_c({\color{black}I}-1)}\int_{0}^{t_{i^*}} e^{-\lambda_c(t_{i^*}-t)} f_T(t) d{\color{black}t} + \int_{t_{i^*}}^{\bar{t}} f_T(t) d{\color{black}t} \bigg)^{{\color{black}I}-1}\hspace{-5mm}. &\label{nofork_ssts} 
	\end{eqnarray}
	\vspace{-4mm}\\
	The equality (a) in \eqref{nofork_ssts} holds since the MMs independently process the computation and transmit the ACK message. Also, equality (b) in \eqref{nofork_ssts} holds since all MMs have identical distributions for $S_i$ and $T_i$. 
	From the equality (c) in \eqref{nofork_ssts}, the computing latency of an MM is non-negative and is greater than $s_{i^*}$, i.e., $s=\max(0,s_{i^*})$. To avoid a forking event, the value of the computing latency has to be greater than $s_{i^*}+t_{i^*}-t$, i.e., $s=\max(0,s_{i^*}+t_{i^*}-t)$. Therefore, the range of the computing latency is  $[\max(s_{i^*}, s_{i^*}+t_{i^*}-t), \infty]$. 
	Moreover, if $s_{i^*}$ and $t_{i^*}$ are given, the PoW computing latency of all MMs other than MM~$i^*$ must be greater than $s_{i^*}$. 
	Therefore, the computing latency of MM $i'$ becomes the conditional probability distribution given by:\vspace{-2mm}
	\begin{eqnarray}
	f_{S}(s|s_{i^*}<S) = \frac{\lambda_c e^{-\lambda_c s}}{1-F_S(s_{i^*})}= \frac{\lambda_c e^{-\lambda_c s}}{e^{-\lambda_c s_{i^*}}}. 
	\end{eqnarray}
\vspace{-4mm}	\\
	Thus, given $s_{i^*}\!$ and $t_{i^*}\!$,  the no-forking probability~yields~\eqref{nofork_ssts}. 
	
	Next, the values of $s_{i^*}$ and $t_{i^*}$ in \eqref{nofork_ssts} follow the probability distributions $f_S(s_{i^*})$ and $f_T(t_{i^*})$, respectively. 
	By integrating \eqref{nofork_ssts} multiplied with $f_S(s_{i^*})$ and $f_T(t_{i^*})$ over the intervals of $s_{i^*}$ and $t_{i^*}$, the no-forking probability is derived as $p_{n}$ in  \eqref{nofork}. 
\end{proof}

\vspace{-8mm}
\bibliographystyle{IEEEtran}


\vspace{-5mm}
\end{document}